\documentclass[10pt,journal,final,compsoc]{IEEEtran}

\ifCLASSOPTIONcompsoc
  \usepackage[nocompress]{cite}
\else
  \usepackage{cite}
\fi
\ifCLASSINFOpdf
\else
\fi
\usepackage{changepage} 
\usepackage{setspace}
\usepackage{datetime}
\usepackage{advdate}
\usepackage{graphicx}
\usepackage{amsmath ,array, amssymb}
\usepackage[colorlinks,
            linkcolor=red,
           anchorcolor=blue,
            citecolor=green
            ]{hyperref} 

\usepackage{tikz}
\def\checkmark{\tikz\fill[scale=0.4](0,.35) -- (.25,0) -- (1,.7) -- (.25,.15) -- cycle;} 
\usepackage{multirow}
\usepackage{float}
\usepackage{footnote}
\newcommand{\etal}{\textit{et al.}} 

\begin{document}
%
\title{A Survey on Knowledge Graph-Based Recommender Systems}
%
%
%
%

\author{Qingyu~Guo,~ 
        Fuzhen~Zhuang,~
        Chuan~Qin,~
        Hengshu~Zhu,~\IEEEmembership{Senior Member,~IEEE,}
        Xing~Xie,~\IEEEmembership{Senior Member,~IEEE,}
        Hui~Xiong,~\IEEEmembership{Fellow,~IEEE,}
        and~Qing~He
\IEEEcompsocitemizethanks{\IEEEcompsocthanksitem Qingyu Guo, Fuzhen Zhuang, Qing He are with Key Lab of Intelligent Information Processing of Chinese Academy of Sciences (CAS), Institute of Computing Technology, CAS, Beijing 100190, China. Qingyu Guo is also with Hong Kong University of
Science and Technology, Clearwater Bay, Kowloon, Hong Kong. Fuzhen Zhuang and Qing He are also with University of Chinese Academy of Sciences, Beijing 100049, China. E-mail: qyguo1996@gmail.com, \{zhuangfuzhen, heqing\}@ict.ac.cn\protect\\
\IEEEcompsocthanksitem Chuan Qin is with University of Science and Technology of China and Baidu Talent Intelligence Center, Baidu Inc. E-mail: chuanqin0426@gmail.com\protect\\
\IEEEcompsocthanksitem Hengshu Zhu and Hui Xiong is with Baidu Talent Intelligence Center, Baidu Inc. Hui Xiong is also with Business Intelligence Lab, Baidu Research. E-mail: \{zhuhengshu, xionghui\}@gmail.com\protect\\
\IEEEcompsocthanksitem Xing Xie is with Microsoft Research Asia, Beijing, China. E-mail: xingx@microsoft.com\protect\\
\IEEEcompsocthanksitem Fuzhen Zhuang is the corresponding author.}
}

\IEEEtitleabstractindextext{%
\begin{abstract}
To solve the information explosion problem and enhance user experience in various online applications, recommender systems have been developed to model users preferences.   
Although numerous efforts have been made toward more personalized recommendations, recommender systems still suffer from several challenges, such as data sparsity and cold start. 
In recent years, generating recommendations with the knowledge graph as side information has attracted considerable interest. Such an approach can not only alleviate the abovementioned issues for a more accurate recommendation, but also provide explanations for recommended items.    
In this paper, we conduct a systematical survey of knowledge graph-based recommender systems. We collect recently published papers in this field and summarize them from two perspectives.   
On the one hand, we investigate the proposed algorithms by focusing on how the papers utilize the knowledge graph for accurate and explainable recommendation. On the other hand, we introduce datasets used in these works.   
Finally, we propose several potential research directions in this field. 
\end{abstract}

}

\maketitle

\IEEEdisplaynontitleabstractindextext

%
\IEEEpeerreviewmaketitle



\section{Introduction}
With the rapid development of the internet, the volume of data has grown exponentially. Because of the overload of information, it is difficult for users to pick out what interests them among a large number of choices. To improve the user experience, recommender systems have been applied for scenarios such as
music recommendation~\cite{hu2018leveraging}, movie recommendation~\cite{Zhang:2016:CKB:2939672.2939673}, and online shopping~\cite{zhao2017meta}.
\par  
The recommendation algorithm is the core element of recommender systems, which are mainly categorized into collaborative filtering (CF)-based recommender systems, content-based recommender systems, and hybrid recommender systems~\cite{adomavicius2005toward}. CF-based recommendation models user preference based on the similarity of users or items from the interaction data, while content-based recommendation utilizes item's content features. 
CF-based recommender systems have been widely applied because they are effective to capture the user preference and can be easily implemented in multiple scenarios, without the efforts of extracting features in content-based recommender systems~\cite{su2009survey,sun2019research}.
However, CF-based recommendation suffers from the data sparsity and cold start problems~\cite{sun2019research}. To address these issues, hybrid recommender systems have been proposed to unify the interaction-level similarity and content-level similarity. In this process, multiple types of side information have been explored, such as item attributes~\cite{sen2009tagommenders,zhen2009tagicofi}, item reviews~\cite{zheng2017joint,xu2018exploiting}, 
and users' social networks~\cite{massa2007trust,jamali2009trustwalker}.
\par
\begin{figure}[!]
	\includegraphics[width=0.48\textwidth]{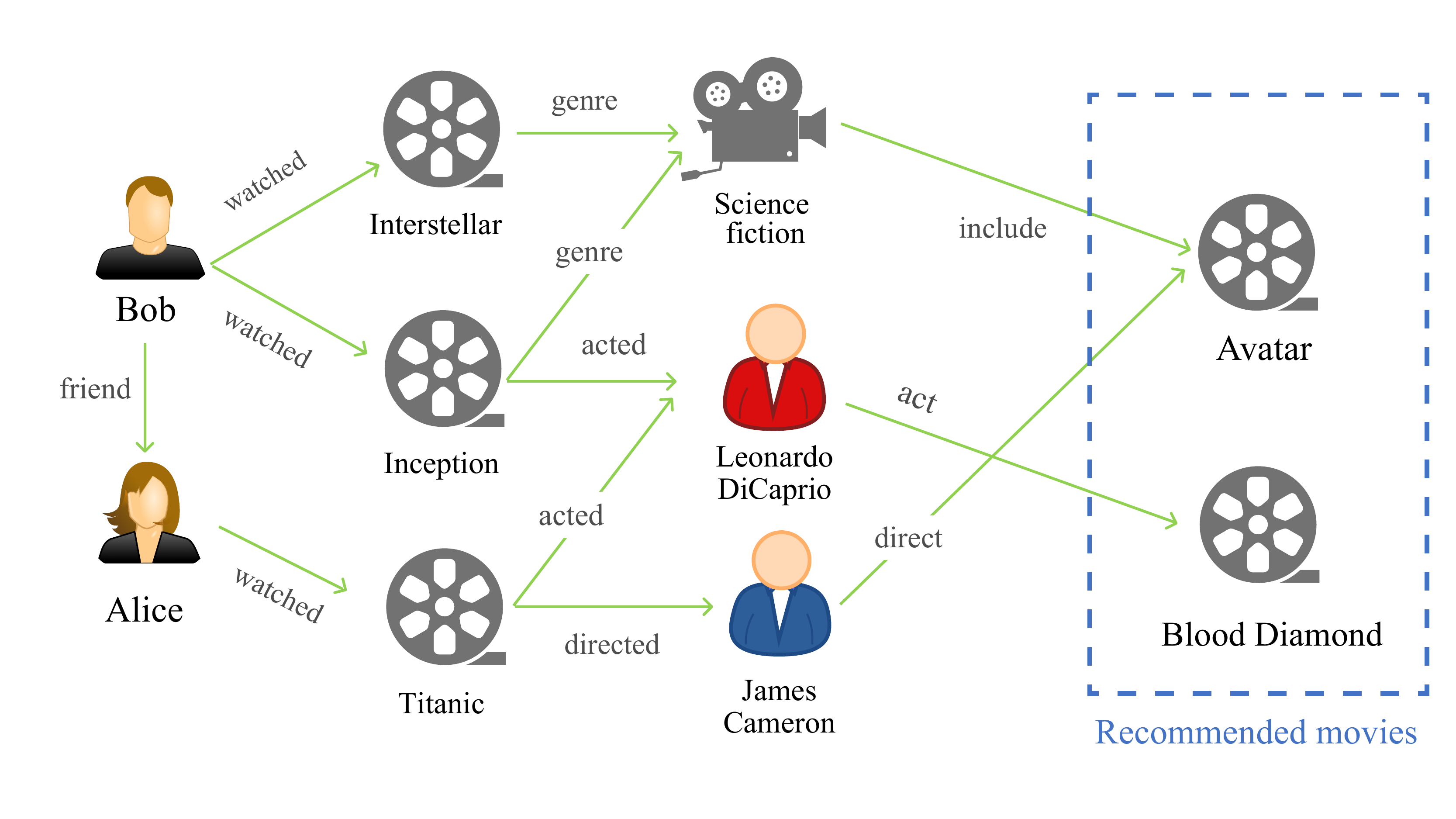}
	\vspace{-3.5ex}
	\caption{An illustration of KG-based recommendation.} 
	\vspace{-3.5ex}
	\label{KG-example}
\end{figure}

   
In recent years, introducing a knowledge graph (KG) into the recommender system as side information has attracted the attention of researchers.
A KG is a heterogeneous graph, where nodes function as entities, and edges represent relations between entities. 
Items and their attributes can be mapped into the KG to understand the mutual relations between items~\cite{Zhang:2016:CKB:2939672.2939673}. Moreover, users and user side information can also be integrated into the KG, which makes relations between users and items, as well as the user preference, can be captured more accurately~\cite{zhang2018learning}.
Figure~\ref{KG-example} is an example of KG-based recommendation, where the movie ``Avatar'' and ``Blood Diamond'' are recommended to Bob. This KG contains users, movies, actors, directors, and genres as entities, while interaction, belonging, acting, directing, and friendship are relations between entities. 
With the KG, movies and users are connected with different latent relations, which helps to improve the precision of recommendation.
Another benefit of KG-based recommender system is the explainability of recommendation results~\cite{wang2018ripplenet}.
In the same example, reasons for recommending these two movies to Bob can be known by following the relation sequences in the user-item graph. For instance, one reason for recommending ``Avatar'' is that ``Avatar'' is the same genre as ``Interstellar'', which was watched by Bob before.
Recently, multiple KGs have been proposed
, such as Freebase~\cite{bollacker2008freebase}, DBpedia~\cite{lehmann2015dbpedia}, YAGO~\cite{suchanek2007yago}, and Google's Knowledge Graph~\cite{singhal2012introducing}, 
which makes it convenient to build KGs for recommendation.
\par   
This survey aims to provide a comprehensive review of the literature utilizing KGs as side information in recommender systems. Throughout our investigation, we discover that existing KG-based recommender systems apply KGs in three ways: the embedding-based method, the path-based method, and the unified method. We illustrate the similarities and differences between these methods in detail. Besides the more accurate recommendation, another benefit of KG-based recommendation is the interpretability. We discuss how different works utilize the KG for explainable recommendation. In addition, based on our survey, we find that KGs serve as side information in multiple scenarios, including the recommendation for movies, books, news, products, points of interest (POIs), music, and social platform. We gather recent works, categorize them by the application, and collect datasets evaluated in these works.
\par
The organization of this survey is as follows: in Section 2, we introduce the foundations of KGs and recommender systems; in Section 3, we present notations and concepts used in this paper; in Section 4 and Section 5, we review KG-based recommender systems from the aspect of approaches and evaluated datasets, respectively; in Section 6, we provide some potential research directions in this field; finally, we conclude this survey in Section 7.

\section{Related Work}
This section introduces the fundamental knowledge and summarizes related work in the domain of KG-based recommendation, including KGs and recommender systems.
\subsection{Knowledge Graphs}
The KG is a practical approach to represent large-scale information from multiple domains~\cite{ehrlinger2016towards}. A common way to describe a KG is to follow the Resource Description Framework (RDF) standard~\cite{gomez2017enterprise}, in which nodes represent entities, while edges in the graph function as relations between entities. Each edge is represented in the form of a triple (head entity, relation, tail entity), also known as a fact in the graph, implying the specific relationship between the head entity and tail entity. For example, (Donald Trump, president\_of, America) indicates the fact that Donald Trump is the president of America. A KG is a heterogeneous network since it contains multiple types of nodes and relations in the graph. Such a graph has strong representation ability since multiple attributes of an entity can be obtained by following different edges in the graph, and high-level relations of entities can be discovered through these relational links.
The concept of knowledge graph was developed in the 1980s~\cite{nurdiati200825}, when KGs were integrated into the framework of expert systems for medical and social sciences. Later, the application was broadened to linguistic and logical domains. 
In 2012, Google introduced the KG into the framework of search, for a better understanding of the query and to make search results more user-friendly~\cite{singhal2012introducing}. 
To date, KGs have been created and applied in multiple scenarios, including search engines, recommender systems, Question Answering system~\cite{huang2019knowledge}, relation detection~\cite{hakkani2014probabilistic}, etc.
\par
We list some popular KGs in Table~\ref{KG_Summary}.
Based on the scope of the knowledge covered, these KGs can be divided into two classes. The first group are cross-domain KGs, such as Freebase~\cite{bollacker2008freebase}, DBpedia~\cite{lehmann2015dbpedia}, YAGO~\cite{suchanek2007yago}, and NELL~\cite{carlson2010coupled}, while the second are domain-specific KGs, like Bio2RDF~\cite{belleau2008bio2rdf}.
Six of the cross-domain KGs are utilized in recommender systems in this survey, and we briefly introduce them as follows:  
Freebase~\cite{bollacker2008freebase} was launched in 2007 by Metaweb and was acquired by Google in 2010.
It contains more than 3 billion facts and almost 50 million entities by 2015~\cite{pellissier2016freebase}.
Though it is a cross-domain KG, around 77\% of its information is in the domain of media~\cite{farber2018knowledge}.
Currently, the data is available at Google's Data Dumps~\cite{freebase:datadumps}.
DBpedia~\cite{lehmann2015dbpedia} is an open community project, which was started by researchers from the Free University of Berlin and Leipzig University, in cooperation with OpenLink Software. The first version was released in 2007 and is updated yearly. The main knowledge is extracted from different language versions of Wikipedia, and DBpedia combines them in a large-scale graph structure. 
YAGO~\cite{suchanek2007yago} (Yet Another Great Ontology) was introduced by the Max Planck Institute in 2007. 
It contains more than 5 million facts, such as people, locations, and organizations. It automatically extracts and unifies knowledge from Wikipedia and multiple sources, including WordNet~\cite{miller1998wordnet} and GeoNames~\cite{GeoNames2006}, then unifies them into an RDF graph.   
Satori~\cite{qian2013understand} is a KG proposed by Microsoft. Similarly to Google's Knowledge Graph, which empowers the Google search engine, Satori has been integrated into the search engine Bing. Though publicly accessible documents about the Satori KG are limited, it is known that Satori consisted of 300 million entities and 800 million relations in 2012~\cite{paulheim2017knowledge}.
CN-DBPedia~\cite{xu2017cn} is the largest Chinese KG. Published by Fudan University in 2015, it has over 16 million entities and over 220 million relations. It automatically extracts knowledge from Baidu Baike, Hudong Baike, and Chinese Wikipedia, then integrates them into a Chinese database. The system is updated continuously with little human effort needed.    
\begin{savenotes}
	\begin{table*}[!]
		\caption{\label{KG_Summary} A collection of commonly used knowledge graphs.}
		\centering
		\begin{tabular}{llll}
			\hline
			KG Name                                     & Domain Type         & Main Knowledge Source  \\ 
			\hline
			YAGO~\cite{suchanek2007yago}                & Cross-Domain        & Wikipedia~\cite{Wikipedia2001}   \\ 
			Freebase~\cite{bollacker2008freebase}       & Cross-Domain             
			& Wikipedia, NNDB~\cite{bettelheim2007nndb}
			FMD~\cite{FMD2000}, MusicBrainz~\cite{Musicbrainz2000}    \\
			DBpedia~\cite{lehmann2015dbpedia}           & Cross-Domain        & Wikipedia       \\
			Satori~\cite{qian2013understand}            & Cross-Domain        & Web Data \\
			CN-DBPedia~\cite{xu2017cn}                  & Cross-Domain        
			& Baidu Baike~\cite{baidubaike2006}, Hudong Baike\cite{hudongbaike2005}, Wikipedia (Chinese)  \\
			NELL~\cite{carlson2010coupled}              & Cross-Domain        & Web Data   \\
			Wikidata~\cite{wikidata2012}                                   & Cross-Domain        & Wikipedia, Freebase\\
			Google's Knowledge Graph~\cite{singhal2012introducing}
			& Cross-Domain        & Web data\\
			Facebook’s Entities Graph~\cite{facebookkg2013}                          & Cross-Domain        & Wikipedia, Facebook data~\cite{facebook2004}        \\
			Bio2RDF~\cite{belleau2008bio2rdf}           & Biological Domain   & Public bioinformatics databases, NCBI’s databases\\
			KnowLife~\cite{ernst2014knowlife}           & Biomedical Domain   & Scientific literature, Web portals\\
			\hline
		\end{tabular}
	\end{table*}
\end{savenotes}    


\subsection{Recommender Systems}
Recommender systems have been applied in many domains, including movies~\cite{Zhang:2016:CKB:2939672.2939673,huang2018improving}, music~\cite{hu2018leveraging,Wang:2019:MFL:3308558.3313411}, POIs~\cite{xi2019modelling,zhao2019go}, news~\cite{Wang:2018:DDK:3178876.3186175,wang2018ripplenet}, education~\cite{huang2019exploring,qin2019duerquiz}, etc. The recommendation task is to recommend one or a series of unobserved items to a given user, and it can be formulated into the following steps. First, the system learns a representation $\textbf{u}_i$ and $\textbf{v}_j$ for the given user $u_i$ and an item $v_j$. Then, it learns a scoring function $f: \textbf{u}_i \times \textbf{v}_j \rightarrow \hat{y}_{i,j} $, which models the preference of $u_i$ for $v_j$. Finally, the recommendation can be generated by sorting the preference scores for items. To learn the user/item representation and the scoring function, there are three main approaches, as described below.
\\
\textbf{$\bullet$ Collaborative Filtering.} CF assumes that users may be interested in items selected by people who share similar interaction records with them. The interaction can either be explicit interaction~\cite{amatriain2009like,jawaheer2010comparison}, like ratings, or implicit interaction~\cite{hu2008collaborative,wang2019setrank}, such as click and view.
To implement CF-based recommendation, interaction data from multiple users and items are required, which further forms the user-item interaction matrix. The CF-based approach contains two main techniques, memory-based CF and model-based CF~\cite{su2009survey}. In detail, memory-based CF first learns the user-user similarity from the user-item interaction data. Then, unobserved items are recommended to a given user based on the interaction records of people similar to the specific user. Alternatively, some models learn the similarity among items, and recommend similar items for a user based on the user's purchase history. 
The model-based CF approach attempts to alleviate the sparsity issue by building an inference model. One common implementation is the latent factor model~\cite{salakhutdinov2008bayesian,zhuang2017representation}, which extracts the latent representation of the user and item from the high dimensional user-item interaction matrix, and then computes the similarity between the user and item with the inner product or other methods.  
\\
\textbf{$\bullet$ Content-based Filtering.} Compared with the CF-based model, which learns the representation of user and item from global user-item interaction data, content-based methods depict the user and item from the content of items.
The assumption of content-based filtering is that users may be interested in items that are similar to their past interacted items. The item representation is obtained by extracting attributes from the item's auxiliary information, including texts, images, etc., while the user representation is based on the features of personal interacted items. The procedure of comparing candidate items with the user profile is essentially matching them with the user's previous records. Therefore, this approach tends to recommend items that are similar to items liked by a user in the past~\cite{lops2011content}.
\\
\textbf{$\bullet$ Hybrid Method.} Hybrid method is to leverage multiple recommendation techniques in order to overcome the limitation of using only one type of method. One major issue of CF-based recommendation is the sparsity of user-item interaction data, which makes it difficult to find similar items or users from the perspective of interaction. A special case for this issue is the cold-start problem, which means the recommendation for new user or item is difficult, since the user-user and item-item similarity cannot be determined without any interaction records. By incorporating content information of users and items, also known as user side information and item side information, into the CF-based framework, better recommendation performance can be achieved~\cite{sun2019research}. 
Some commonly used item side information include item attributes~\cite{sen2009tagommenders, zhen2009tagicofi,ziegler2004taxonomy,ziegler2004taxonomy}, like brands, categories; item multimedia information, like textual description~\cite{han2019adaptive}, image features~\cite{chu2017hybrid}, audio signals~\cite{liang2015content}, video features~\cite{chen2017attentive}; and item reviews~\cite{zheng2017joint,xu2018exploiting}. Common options for user side information involve user's demographic information~\cite{gantner2010learning}, including occupation, gender, and hobbies; and user network~\cite{massa2007trust,jamali2009trustwalker}. In this survey, KG-based recommender systems leverage the KG as the side information, combining the CF-based technique for more accurate recommendation. 
\\ 

\section{Overview}
\label{section3:overview}
Before delving into the state-of-the-art approaches exploiting KGs as side information for recommendation, we first present notations and concepts used in the paper to eliminate misunderstanding. For convenience, we list some symbols and their descriptions in Table~\ref{Notation}.
\begin{table*}[!]
	\caption{\label{Notation} Notations used in this paper.}
	\centering
	\begin{tabular}{ll}
		\hline
		Notations & Descriptions \\ \hline
		$u_i$         & User $i$             \\
		$v_j$         & Item $j$             \\
		$e_k$         & Entity $k$ in the knowledge graph \\ 
		$r_k$         & Relation between two entities ($e_i$, $e_j$) in the knowledge graph          \\
		$\hat{y}_{i, j}$         & Predicted user $u_i$'s preference for item $v_j$           \\
		$\textbf{u}_i \in \mathbb{R}^{d \times 1} $         & Latent vector of user $u_i$             \\
		$\textbf{v}_j \in \mathbb{R}^{d \times 1}$         & Latent vector of item $v_j$             \\
		$\textbf{e}_k \in \mathbb{R}^{d \times 1}$         & Latent vector of entity $e_k$ in the KG            \\
		$\textbf{r}_k \in \mathbb{R}^{d \times 1}$         & Latent vector of relation $r_k$ in the KG            \\
		$\mathcal{U}=\left\{u_1, u_2, \cdots, u_m\right\}$         & User set             \\
		$\mathcal{V}=\left\{v_1, v_2, \cdots, v_n\right\}$         & Item set             \\
		$\textbf{U} \in \mathbb{R}^{d \times m} $         & Latent vector of the user set             \\
		$\textbf{V} \in \mathbb{R}^{d \times n}$         & Latent vector of the item set             \\
		$R \in \mathbb{R}^{m \times n}$         & User-Item Interaction matrix             \\
		$p_k$         & One path $k$ to connect two entities ($e_i$, $e_j$) in the knowledge graph        \\
		$\mathcal{P}(e_i, e_j)=\left\{p_1, p_2, \cdots, p_s\right\}$         & Path set between entity pair ($e_i$, $e_j$)            \\
		$\Phi$ & Nonlinear Transformation \\
		$\odot$ & Element-wise Product\\
		$\oplus$ & Vector concatenation operation \\
		\hline
	\end{tabular}
\end{table*}
\\
\textbf{$\bullet$ Heterogeneous Information Network (HIN).} 
A HIN is a directed graph \(G=(V, E)\) with an entity type mapping function \(\phi: V \rightarrow \mathcal{A}\) and a link type mapping
function \(\psi: E \rightarrow \mathcal{R} .\) Each entity \(v \in V\) belongs to an entity
type \(\phi(v) \in \mathcal{A},\) and each link \(e \in E\) belongs to a relation
type \(\psi(e) \in \mathcal{R} .\) In addition, the number of entity types \(|\mathcal{A}|>1\) and/or the number of relation types \(|\mathcal{R}|>1\).
\\
\textbf{$\bullet$ Knowledge Graph (KG).}
A KG \(\mathcal{G}_{k n o w}=(V, E)\) is a directed graph whose nodes are entities and edges are subject-property-object triple facts. Each edge of the form (\textit{head entity, relation, tail entity}) (denoted as \(<e_h, r, e_t>\)) indicates a relationship of $r$ from entity $e_h$ to entity $e_t$. It can be regarded as an instance of a HIN.
\\
\textbf{$\bullet$ Meta-path.} 
A meta-path 
\(\mathcal{P}=A_{0} \stackrel{R_{1}}{\longrightarrow}\)
\(A_{1} \stackrel{R_{2}}{\longrightarrow} \cdots \stackrel{R_{k}}{\longrightarrow} A_{k}\)
is a path defined on the graph of network schema \(G_{T}=(\mathcal{A}, \mathcal{R}),\) which defines a new composite relation \(R_{1} R_{2} \cdots R_{k}\) between type \(A_{0}\) and \(A_{k},\) where \(A_{i} \in \mathcal{A}\) and \(R_{i} \in \mathcal{R}\) for \(i=0, \cdots, k.\) 
It is a relation sequence connecting object pairs in a HIN, which can be used to extract connectivity features in the graph. 
\\
\textbf{$\bullet$ Meta-graph.} Similar to a meta-path, a meta-graph is another meta-structure that connects two entities in a HIN. The difference is that a meta-path only defines one relation sequence, while a meta-graph is a combination of different meta-paths~\cite{fang2016semantic}. Compared with a meta-path, a meta-graph can contain more expressive structural information between entities in the graph.  
\\
\textbf{$\bullet$ Knowledge Graph Embedding (KGE).} 
KGE is to embed a KG \(\mathcal{G}_{k n o w}=(V, E)\) into a low dimensional space~\cite{cai2018comprehensive}. After the embedding procedure, each graph component, including the entity and the relation, is represented with a $d$-dimensional vector. The low dimensional embedding still preserves the inherent property of the graph, which can be quantified by semantic meaning or high-order proximity in the graph.
\\
\textbf{$\bullet$ User Feedback.} 
With \(m\) users \(\mathcal{U}=\left\{u_{1}, \cdots, u_{m}\right\}\) and \(n\) items \(\mathcal{V}=\left\{v_{1}, \cdots, v_{n}\right\}\), we define the binary user feedback matrix \(R \in \mathbb{R}^{m \times n}\) as follows:
\begin{equation*}
R_{i j}=\left\{\begin{array}{ll}{1,} & {\text { if }\left(u_{i}, v_{j}\right) \text { interaction is observed; }} \\ {0,} & {\text { otherwise. }}\end{array}\right.
\end{equation*}
Note that a value of 1 for $R_{ij}$ indicates there is an implicit interaction between user $u_i$ and item $v_j$, such as behaviors of clicking, watching, browsing, etc. Such an implicit interaction does not necessarily imply $u_i$'s preference over $v_j$. Unless otherwise stated, the user feedback used in this paper means the implicit feedback. However, in some specific scenarios, explicit feedback to show the user's preference can also be available. For example, in movie recommendation, a user explicitly rates a movie in the score range of one to five. Some papers have extracted the data of score ratings of five to indicate the user's preference in such a case.
\\
\textbf{$\bullet$ $H$-hop Neighbor.} 
Nodes in the graph can be connected with a multi-hop relation path: $e_{0} \stackrel{r_{1}}{\longrightarrow} e_{1} \stackrel{r_{2}}{\longrightarrow} \cdots \stackrel{r_{H}}{\longrightarrow} e_{H}$, in this case, $e_{H}$ is the H-hop neighbor of $e_{0}$, which can be represented as $e_{H} \in \mathcal{N}_{e_{0}}^H$. Note that $\mathcal{N}_{e_{0}}^0$ is $e_0$ itself.
\\
\textbf{$\bullet$ Relevant Entity.}
Given the interaction matrix $R$ and the knowledge graph $\mathcal{G}_{k n o w}$, the set of $k$-hop relevant entities for user u can be represented as 
\begin{equation*}
\begin{aligned}
&\mathcal{E}_{u}^{k}=\left\{e_t |(e_h, r, e_t) \in \mathcal{G} \text { and } e_h \in \mathcal{E}_{u}^{k-1}\right\},
\\ & k=1,2, \cdots, H.
\end{aligned}
\end{equation*}
where \(\mathcal{E}_{u}^{0}=\left\{u | R_{u v}=1\right\}\) is the set of the user's historical interacted items. 
\\
\textbf{$\bullet$ User Ripple Set.} 
The ripple set of a user is defined as the knowledge triples with the head entities being $(k-1)$-hop relevant entities $\mathcal{E}_{u}^{k-1}$, 
\begin{equation*}
\begin{aligned}
&\mathcal{S}_{u}^{k}=\left\{(e_h, r, e_t) |(e_h, r, e_t) \in \mathcal{G} \text { and } e_h \in \mathcal{E}_{u}^{k-1}\right\}, 
\\ &k=1,2, \cdots, H.
\end{aligned}
\end{equation*}
\\ 
\textbf{$\bullet$ Entity Ripple Set.} 
The ripple set of an entity $e \in \mathcal{G}$ is defined as 
\begin{equation*}
\begin{aligned}
&\mathcal{S}_{e}^{k}=\left\{(e_h, r, e_t) |(e_h, r, e_t) \in \mathcal{G} \text { and } e_h \in \mathcal{N}_e^{k-1} \right\}, 
\\ &k=1,2, \cdots, H.
\end{aligned}
\end{equation*}

\section{Methods of Recommender Systems with Knowledge Graphs}
In this section, we collect papers related to KG-based recommender systems. Based on how these works utilize the KG information, we group them into three categories: embedding-based methods, path-based methods, and unified methods. We will introduce how different methods leverage KGs to improve the recommendation results. To facilitate readers checking the literature, we summarize and organize these papers in Table~\ref{papers}, which lists their publication information, the approach to utilize a KG for recommendation, and the techniques adopted in these works.
\begin{table*}[!]
	\caption{\label{papers} Table of collected papers. In the table, `Emb.' stands for embedding-based Method, `Uni.' stands for unified method, `Att.' stands for attention mechanism, `RL' stands for reinforcement learning, `AE' stands for autoencoder, and `MF' stands for matrix factorization.}
	\centering
	\begin{tabular}{llllll|llllllll}
		\hline
		\multirow{2}{*}{\textbf{Method}} & \multirow{2}{*}{\textbf{Venue}} & \multirow{2}{*}{\textbf{Year}} & \multicolumn{3}{c}{\textbf{KG Usage Type}}
		& \multicolumn{8}{c}{\textbf{Framework}}    \\ \cline{4-6} \cline{7-14}  
		&                                 &                                & \textbf{Emb.} & \textbf{Path} & \textbf{Uni.} 
		& CNN       & RNN       & Att.      & GNN       & GAN       & RL        & AE        & MF 
		\\ \hline
		CKE\cite{Zhang:2016:CKB:2939672.2939673}                            & KDD                             & 2016                           & \checkmark            &               &                
		&           &           &           &           &           &           &\checkmark & \\
		entity2rec\cite{palumbo2017entity2rec}                       & RecSys                          & 2017                           & \checkmark            &               &                
		\\
		ECFKG\cite{ai2018learning}                            & Algorithms                      & 2018                           & \checkmark            &               &                
		\\
		SHINE\cite{wang2018shine}                            & WSDM                            & 2018                           & \checkmark            &               &                
		&           &           &           &           &           &           &\checkmark &\\
		DKN\cite{Wang:2018:DDK:3178876.3186175}                              & WWW                             & 2018                           & \checkmark            &               &                
		&\checkmark &           &\checkmark &           &           &           &           &\\
		KSR\cite{huang2018improving}                              & SIGIR                           & 2018                           & \checkmark            &               &                
		&           &\checkmark &\checkmark &           &           &           &           &\\
		CFKG\cite{zhang2018learning}                             & SIGIR                           & 2018                           & \checkmark            &               &                
		\\
		KTGAN\cite{yang2018knowledge}                            & ICDM                            & 2018                           & \checkmark            &               &                
		&           &           &           &           &\checkmark &           &           &\\
		KTUP\cite{Cao:2019:UKG:3308558.3313705}                             & WWW                             & 2019                           & \checkmark            &               &                
		\\
		MKR\cite{Wang:2019:MFL:3308558.3313411}                              & WWW                             & 2019                           & \checkmark            &               &                
		&           &           &\checkmark &           &           &           &           &\\
		DKFM\cite{dadoun2019location}                             & WWW                             & 2019                           & \checkmark            &               &                
		\\
		SED\cite{joseph2019content}                              & WWW                             & 2019                           & \checkmark            &               &                
		\\
		RCF\cite{Xin:2019:RCF:3331184.3331188}                              & SIGIR                           & 2019                           & \checkmark            &               &                
		&           &           &\checkmark &           &           &           &           &\\
		BEM\cite{ye2019bayes}                              & CIKM                            & 2019                           & \checkmark            &               &                
		\\
		Hete-MF\cite{yu2013collaborative}                          & IJCAI                           & 2013                           &              & \checkmark             &
		&           &           &           &           &           &           &           &\checkmark                \\
		HeteRec\cite{yu2013recommendation}                          & RecSys                          & 2013                           &              & \checkmark             &                
		&           &           &           &           &           &           &           &\checkmark \\
		HeteRec\_p\cite{yu2014personalized}                       & WSDM                            & 2014                           &              & \checkmark             &                
		&           &           &           &           &           &           &           &\checkmark \\
		Hete-CF\cite{luo2014hete}                          & ICDM                            & 2014                           &              & \checkmark             &                
		&           &           &           &           &           &           &           &\checkmark\\
		SemRec\cite{shi2015semantic}                           & CIKM                            & 2015                           &              & \checkmark             &                
		&           &           &           &           &           &           &           &\checkmark\\
		ProPPR\cite{catherine2016personalized}                           & RecSys                          & 2016                           &              & \checkmark             &                
		&           &           &           &           &           &           &           &\checkmark\\
		FMG\cite{zhao2017meta}                              & KDD                             & 2017                           &              & \checkmark             &                
		&           &           &           &           &           &           &           &\checkmark\\
		MCRec\cite{hu2018leveraging}                            & KDD                             & 2018                           &              & \checkmark             &                
		&\checkmark &           &\checkmark &           &           &           &           &\checkmark\\
		RKGE\cite{Sun:2018:RKG:3240323.3240361}                             & RecSys                          & 2018                           &              & \checkmark             &                
		&           &\checkmark &\checkmark &           &           &           &           &           \\
		HERec\cite{shi2018heterogeneous}                            & TKDE                            & 2019                           &              & \checkmark             &                
		&           &           &           &           &           &           &           &\checkmark\\
		KPRN\cite{wang2019explainable}                             & AAAI                            & 2019                           &              & \checkmark             &                
		&           &\checkmark &\checkmark &           &           &           &           &           \\
		RuleRec\cite{ma2019jointly}                          & WWW                             & 2019                           &              & \checkmark             &                
		&           &           &           &           &           &           &           &\checkmark\\
		PGPR\cite{xian2019reinforcement}                             & SIGIR                           & 2019                           &              & \checkmark             &                
		&           &           &           &           &           &\checkmark &           &           \\
		EIUM\cite{huang2019explainable}                             & MM                              & 2019                           &              &    \checkmark          &                
		&\checkmark &           &\checkmark &           &           &           &           &\\
		Ekar\cite{song2019explainable}                             & arXiv                           & 2019                           &              & \checkmark             &                
		&           &           &           &           &           &\checkmark &           &           \\
		RippleNet\cite{wang2018ripplenet}                        & CIKM                            & 2018                           &              &               & \checkmark              
		&           &           &\checkmark & &           &           &           &\\
		{RippleNet-agg}\cite{wang2019exploring}                    & TOIS                            & 2019                           &              &               & \checkmark              
		&           &           &\checkmark &\checkmark &           &           &           &\\
		{KGCN}\cite{Wang:2019:KGC:3308558.3313417}                             & WWW                             & 2019                           &              &               & \checkmark              
		&           &           &\checkmark &\checkmark &           &           &           &\\
		{KGAT}\cite{Wang:2019:KKG:3292500.3330989}                             & KDD                             & 2019                           &              &               & \checkmark              
		&           &           &\checkmark &\checkmark &           &           &           &\\
		{KGCN-LS}\cite{Wang:2019:KGN:3292500.3330836}                          & KDD                             & 2019                           &              &               & \checkmark              
		&           &           &\checkmark &\checkmark &           &           &           &\\
		{AKUPM}\cite{tang2019akupm}                            & KDD                             & 2019                           &              &               & \checkmark              
		&           &           &\checkmark &           &           &           &           &\\
		{KNI}\cite{qu2019end}                              & KDD                             & 2019                           &              &               & \checkmark              
		&           &           &\checkmark &\checkmark &           &           &           &\\
		{IntentGC}\cite{zhao2019intentgc}                         & KDD                             & 2019                           &              &               & \checkmark              
		&           &           &           &\checkmark &           &           &           &\\
		{RCoLM}\cite{li2019unifying}                            & IEEE Access                     & 2019                           &              &               & \checkmark              
		&           &           &\checkmark &           &           &           &           &\\
		{AKGE}\cite{sha2019attentive}                             & arXiv                           & 2019                           &              &               & \checkmark              
		&           &           &\checkmark &\checkmark &           &           &           &\\ \hline
	\end{tabular}
\end{table*}
\par
Explainable recommendation has been another hot research topic in recent years. It is helpful for users to adopt the suggestions generated by the recommender system if appropriate explanations are provided to them~\cite{zhang2018explainable}. Compared with traditional recommender systems, KG-based recommendation makes the reasoning process available. In this section, we will also show how different works leverage KGs for explainable recommendation.  
\subsection{Embedding-based Methods}
\label{section4-1:embedding}
The embedding-based methods generally use the information from the KG directly to enrich the representation of items or users. In order to exploit the KG information, knowledge graph embedding (KGE) algorithms need to be applied to encode the KG into low-rank embedding. KGE algorithms can be divided into two classes~\cite{wang2017knowledge}: translation distance models, such as TransE~\cite{bordes2013translating}, TransH~\cite{wang2014knowledge}, TransR~\cite{lin2015learning}, TransD~\cite{ji2015knowledge}, etc., and semantic matching models, such as DistMult~\cite{yang2014embedding}. \par
Based on whether users are included in the KG, embedding-based methods can be divided into two classes.  
In the first type of method, KGs are constructed with items and their related attributes, which are extracted from the dataset or external knowledge bases. We name such a graph as the item graph. Note that users are not included in such an item graph. Papers following this strategy leverage the knowledge graph embedding (KGE) algorithms to encode the graph for a more comprehensive representation of items, and then integrate the item side information into the recommendation framework. 
The general idea can be illustrated as follows.
The latent vector $\textbf{v}_j$ of each item $v_j$ is obtained by aggregating information from multiple sources, such as the KG, the user-item interaction matrix, item's content, and item's attributes.  
The latent vector $\textbf{u}_i$ of each user $u_i$ can either be extracted from the user-item interaction matrix, or the combination of interacted items' embedding.
Then, the probability of $u_i$ selecting $v_j$ can be calculated with 
\begin{equation}
\hat{y}_{i, j} = f(\textbf{u}_i,\textbf{v}_j),
\label{equation:emb-predict}
\end{equation}
where $f(\cdot)$ refers to a function to map the embedding of the user and item into a preference score, which can be the inner product, DNN, etc. In the recommendation stage, results will be generated in descending order of the preference score $\hat{y}_{i, j}$. 
\par
For instance, Zhang \etal~\cite{Zhang:2016:CKB:2939672.2939673} proposed CKE, which unifies various types of side information in the CF framework. They fed the item's structural knowledge (item's attributes represented with knowledge graph) and content (textual and visual) knowledge into a knowledge base embedding module. The latent vector of the item's structural knowledge $\textbf{x}_j$ is encoded with the TransR algorithm, while the textual feature $\textbf{z}_{t,j}$ and the visual feature $\textbf{z}_{v,j}$ are extracted with the autoencoder architecture. Then these representations are aggregated along with the offset vector $\boldsymbol{\eta}_j$ extracted from the user-item interaction matrix. The final representation of each item $v_j$ can be written as
\begin{equation}
\mathbf{v}_{j}=\boldsymbol{\eta}_{j}+\mathbf{x}_{j}+\mathbf{z}_{t, j}+\mathbf{z}_{v, j}.
\end{equation}
After obtaining the latent vector $\textbf{u}_i$ of the user $u_i$, the preference score $\hat{y}_{i, j}$ is obtained via the inner product $\textbf{u}_i^T\textbf{v}_j$. Finally, in the prediction stage, items are recommended to $u_i$ by the following ranking criteria:
\begin{small}
	\begin{equation}
	v_{j_{1}}>v_{j_{2}}>\cdots>v_{j_{n}} \rightarrow \mathbf{u}_{i}^{T} \mathbf{v}_{j_{1}}>\mathbf{u}_{i}^{T} \mathbf{v}_{j_{2}}>\cdots>\mathbf{u}_{i}^{T} \mathbf{v}_{j_{n}}.
	\end{equation}
\end{small}
Experiments show that incorporating structural knowledge can boost the performance of recommendation.
\par
Wang \etal~\cite{Wang:2018:DDK:3178876.3186175} proposed DKN for news recommendation. It models the news by combining the textual embedding of sentences learned with Kim CNN~\cite{kim2014convolutional} and the knowledge-level embedding of entities in news content via TransD. With the incorporation of a KG for entities, high-level semantic relations of news can be depicted in the final embedding $\textbf{v}_j$ of news $v_j$. In order to capture the user's dynamic interest in news, the representation of $u_i$ is learned by aggregating the embedding of historical clicked news $\{\textbf{v}_1,\textbf{v}_2,\cdots,\textbf{v}_{N_i}\}$ with an attention mechanism. The attention weight for each news $v_k(k=1,2,\cdots,N_i)$ in the clicked news set is calculated via
\begin{equation}
s_{v_{k}, v_{j}}=\frac{\exp \left(g\left(\mathbf{v}_k, \mathbf{v}_j\right)\right)}{\sum_{k=1}^{N_{i}} \exp \left(g\left(\mathbf{v}_k, \mathbf{v}_j\right)\right)},
\end{equation}
where $g(\cdot)$ is a DNN layer, $v_j$ is the candidate news.
Then, the final user embedding $\mathbf{u}_i$ is calculated via the weighted sum of clicked news embeddings:
\begin{equation}
\mathbf{u}_i=\sum_{k=1}^{N_{i}} s_{v_{k}, v_{j}} \mathbf{v}_k.
\end{equation} 
Finally, user's preference for candidate news $v_j$ can be calculated with Equation~\ref{equation:emb-predict}, where $f(\cdot)$ is a DNN layer. 
Huang \etal~\cite{huang2018improving} proposed the KSR framework for sequential recommendation. KSR uses a GRU network with a knowledge-enhanced key-value memory network (KV-MN) to model comprehensive user preference from the sequential interaction. The GRU network captures the user's sequential preference, while the KV-MN module utilizes knowledge base information (learned with TransE) to model the user's attribute-level preference. In this way, fine-grained user preference can be captured for recommendation.
In detail, at time $t$, the latent vector of $u_i$ is represented as $\textbf{u}_i^t=\textbf{h}_i^t \oplus \textbf{m}_i^t$, where $\textbf{h}_i^t$ and $\textbf{m}_i^t$ stands for the representation of user's interaction-level preference and attribute-level preference, respectively. The latent vector of $v_j$ is represented as $\textbf{v}_j=\textbf{q}_j \oplus \textbf{e}_j \cdot \textbf{u}_i^t$, where $\textbf{q}_j$ is the item embedding in the GRU network, and $\textbf{e}_j$ is the item embedding in the KG. After transforming $\textbf{u}_i^t$ and $\textbf{v}_j$ to the same dimension, the user's preference for items is ranked with the score obtained from Equation~\ref{equation:emb-predict}, where $f(\cdot)$ is the inner product. 
\par
The other type of embedding-based method directly builds a user-item graph, where users, items, and their related attributes function as nodes. In the user-item graph, both attribute-level relation (brand, category, etc) and user-related relations (co-buy, co-view, etc.) serve as edges.
After obtaining the embeddings of entities in the graph, the user's preference can be calculated with Equation~\ref{equation:emb-predict}, or by further considering the relation embedding in the graph via
\begin{equation}
\hat{y}_{i, j} = f(\textbf{u}_i,\textbf{v}_j,\textbf{r}),
\end{equation}
where $f(\cdot)$ maps the user representation $\textbf{u}_i$, the item representation $\textbf{v}_j$, as well as the relation embedding $\textbf{r}$ into a scalar.   
\par
Zhang \etal~\cite{zhang2018learning} proposed CFKG, which constructs a user-item KG. In this user-item graph, user behaviors (purchase, mention) are regarded as one relation type between entities, and multiple types of item side information (review, brand, category, bought-together, etc.) are included.
To learn the embedding of entities and relations in the graph, the model defines a metric function $d(\cdot)$ to measure the distance between two entities according to a given relation. In the recommendation phase, the system will rank candidate items $j$ in an ascending order of the distance between $u_i$ and $v_j$  
\begin{equation}
d\left(\textbf{u}_{i}+\textbf{r}_{buy},\textbf{v}_{j}\right),
\end{equation}
where $\textbf{r}_{buy}$ is the learned embedding for the relation type `buy'. A smaller distance between $u_i$ and $v_j$ measured by the `buy' relation refers to a higher preference score $\hat{y}_{i, j}$.
\par
Wang \etal~\cite{wang2018shine} proposed SHINE, which takes the celebrity recommendation task as the sentiment link prediction task between entities in the graph. In detail, SHINE builds a sentiment network $G_s$ for users and targets (celebrities), and utilizes their social network $G_r$ and profile information network $G_p$ as side information. These three networks are embedded with the auto-encoder technique, and are then aggregated as the representation of the user and target. Finally, the recommendation can be generated by following Equation~\ref{KG_Summary}, where $f(\cdot)$ is a DNN layer.
Dadoun \etal~\cite{dadoun2019location} proposed DKFM for POI recommendation. DKFM applies TransE over a city KG to enrich the representation of the destination, which shows improvement in the performance of POI recommendation.
\par
Previous works generally directly utilize the raw latent vector of structural knowledge learned with the KGE technique for recommendation. Recently, some papers have tried to improve the recommendation performance by refining the learned entity/relation representation.
For instance, Yang \etal~\cite{yang2018knowledge} introduced a GAN-based model, KTGAN, for movie recommendation. 
In the first phase, KTGAN learns the knowledge embedding $\textbf{v}_j^k$ for movie $v_j$ by incorporating the Metapath2Vec model~\cite{dong2017metapath2vec} on the movie's KG, and the tag embedding $\textbf{v}_j^t$ with the Word2Vec model~\cite{mikolov2013efficient} on movie's attributes. The initial latent vector of movie $v_j$ is represented as $\textbf{v}_j^{initial}=\textbf{v}_j^k \oplus \textbf{v}_j^t$. Similarly, the initial latent vector of user $u_i$ is represented as $\textbf{u}_i^{initial}=\textbf{u}_i^k \oplus \textbf{u}_i^t$, where $\textbf{u}_i^k$ is the average of knowledge embeddings of $u_i$'s favored movies, and $\textbf{u}_i^t$ is $u_i$'s tag embedding.
Then, a generator $G$ and a discriminator $D$ are proposed to refine initial representations of users and items. The generator $G$ tries to generate relevant (favorite) movies for user $u_i$ according the score function $p_{\theta}(v_j|u_i,r)$, where $r$ denotes the relevance between $u_i$ and $v_j$. During the training process, $G$ aims to let $p_{\theta}(v_j|u_i,r)$ approximate $u_i$'s true favorite movie distribution $p_{true}(v_j|u_i,r)$, so that $G$ can select relevant user-movie pairs. The discriminator $D$ is a binary classifier to distinguish relevant user-movie pairs and irrelevant pairs according to the learned score function $f_{\phi}(u_i,v_j)$. The objective function of the GAN module is written as, 
\begin{equation}
\begin{aligned}
\mathcal{L}=\min_{\theta}\max_{\phi} &\sum_{i=1}^{M} \{
\mathbb{E}_{v_j \sim p_{\text {true }}\left(v_j | u_{i}, r\right)}\left[\log P\left(v_j | u_{i}\right)\right] \\
+ & \mathbb{E}_{v_j \sim p_{\theta}\left(v_j | u_{i}, r\right)}\left[\log \left(1-P\left(v_j | u_{i}\right)\right)\right]\},
\end{aligned}
\end{equation}  
where  $
P(v_j | u_i)=\frac{1}{1+\exp \left(-f_{\phi}(u_i, v_j)\right)}
$ stands for the probability of movie $v_j$ being preferred by user $u_i$. After the adversarial training, optimal representations of $u_i$ and $v_j$ are learned and movies can be ranked with $G$'s score function $p_{\theta}(v_j|u_i,r)$.
Later, Ye \etal~\cite{ye2019bayes} proposed BEM, which uses two types of graphs for items, the knowledge-related graph (containing item attributes information, like brand, category, etc.) and behavior graph (containing item interaction-related information, including co-buy, co-rate, co-add to cart) for recommendation. BEM first learns the initial embeddings from the knowledge-related graph and the behavior graph with the TransE model and a GNN-based model, respectively. Then, BEM applies a Bayesian framework to refine these two types of embeddings mutually. Recommendation can be generated by finding the closest items of the interacted items in the behavior graph, which are measured by the relation of `co-buy' or `co-click'. 
\par
Another trend is to adopt the strategy of multi-task learning, to jointly learn the recommendation task with the guidance of the KG-related task. Generally, in the recommendation task, a function ${f}(\textbf{u}_i,\textbf{v}_j)$ is learned from the user-item interaction matrix to contrast the observed interaction pair $(u_i,v_j)$ and unobserved interaction pair $(u_i,v_j\prime)$; in the KG-related task, another function $g(\textbf{e}_h,\textbf{r},\textbf{e}_t)$ is learned to determine whether $(e_h,r,e_t)$ is a valid triplet in the KG. These two parts are connected with the following objective function,
\begin{equation}
\mathcal{L}=\mathcal{L}_{rec}+\lambda \mathcal{L}_{KG},
\end{equation} 
where $\mathcal{L}_{rec}$ is the loss function for recommendation, $\mathcal{L}_{KG}$ is the loss function for the KG related task, and $\lambda$ is the hyperparameter to balance the two tasks. A general motivation for the multi-task learning is that item embeddings in the recommendation module share features with the associated entity embeddings in the KG. 
\par
Cao \etal~\cite{Cao:2019:UKG:3308558.3313705} proposed KTUP to jointly learn the task of recommendation and knowledge graph completion.  
In the recommendation module, the loss function is defined as
\begin{equation}
\mathcal{L}_{rec}=\sum_{(u, v,v') \in {R}}-\log \sigma\left[f\left(\textbf{u}, \textbf{v}^{\prime}, \textbf{p}^{\prime}\right)-f(\textbf{u},\textbf{v}, \textbf{p})\right],
\end{equation}
where $(u,v)$ is the observed user-item pair in the user-item interaction matrix ($R_{uv}=1$); $(u,v\prime)$ denotes the unobserved user-item pair ($R_{uv\prime}=0$); $\textbf{p}$ denotes the latent vector of user's preference for the given item; $f(\cdot)$ is the proposed translation-based model, TUP, to model the correctness of such a user-item pair; and $\sigma$ is the sigmoid function.
For the KG completion module, a hinge loss is adopted,
\begin{footnotesize}
	\begin{equation}
	\mathcal{L}_{KG}=\sum_{\left(e_h, r, e_t\right) \in  \mathcal{G}} \sum_{\left(e_h^{\prime}, r^{\prime}, e_t^{\prime}\right) \in \mathcal{G}^{-}}\left[g\left(\textbf{e}_h, \textbf{r}, \textbf{e}_t\right)+\gamma-g\left(\textbf{e}_h^{\prime}, \textbf{r}^{\prime}, \textbf{e}_t^{\prime}\right)\right]_{+},
	\end{equation}    
\end{footnotesize}
where $\mathcal{G}^{-}$ is constructed by replacing $e_h$ or $e_t$ in the valid triplet $(e_h,r,e_t) \in \mathcal{G}$; $g(\cdot)$ is the TransH model, and a lower $g(\textbf{e}_h,\textbf{r},\textbf{e}_t)$ value infers a higher correctness of such a triplet; $[\cdot]_{+} \triangleq \max (0, \cdot)$; and $\gamma$ is the margin between correct triplets and incorrect triplets. The recommendation module is to mine the preference relation between user $u$ and item $v$, while the knowledge completion task is to mine the relation among items in the KG. 
The bridge between these two modules is that items can be aligned with corresponding entities in the KG, and the user's preference is related with relations among entities in the KG. Hence, embeddings of items and preferences can be enriched by transferring knowledge of entities, relations and preference in each module under the framework of KTUP.
Meanwhile, Wang \etal~\cite{Wang:2019:MFL:3308558.3313411} proposed MKR, which consists of a recommendation module and a KGE module. The former learns latent representation for users and items, while the latter learns representation for item associated entities with the semantic matching KGE model. These two parts are connected with a cross \& compress unit to transfer knowledge and share regularization of items in the recommendation module and entities in the KG.
Xin \etal~\cite{Xin:2019:RCF:3331184.3331188} proposed RCF, which introduces a hierarchical description of items, including both the relation type embedding and relation value embedding. RCF utilizes the DistMult model for KGE to preserve the relational structure between items. Then, it models the user's type-level preference and value-level preference separately with the attention mechanism. With the jointly training of recommendation module and the KG relation modeling module, decent recommendations can be made.
\\
\textbf{Summary for Embedding-based Methods.} Most embedding-based methods~\cite{Zhang:2016:CKB:2939672.2939673,Wang:2018:DDK:3178876.3186175,huang2018improving,joseph2019content,yang2018knowledge,ye2019bayes,Cao:2019:UKG:3308558.3313705,Wang:2019:MFL:3308558.3313411,Xin:2019:RCF:3331184.3331188} build KGs with multiple types of item side information to enrich the representation of items, and such information can be used to model the user representation more precisely. Some models~\cite{ai2018learning,zhang2018learning,wang2018shine,palumbo2017entity2rec,dadoun2019location} build user-item graphs by introducing users into the graph, which can directly model the user preference. Entity embedding is the core of embedding-based methods, and some papers refine the embedding with GAN~\cite{yang2018knowledge} or BEM~\cite{ye2019bayes} for better recommendation. Embedding-based methods leverage the information in the graph structure intrinsically. Papers~\cite{Cao:2019:UKG:3308558.3313705,Wang:2019:MFL:3308558.3313411,Xin:2019:RCF:3331184.3331188} apply the strategy of multi-task learning to jointly train the recommendation module along with the graph-related task to improve the quality of recommendation. 

\subsection{Path-based Methods}
\label{section4-2:path}
Path-based methods build a user-item graph and leverage the connectivity patterns of the entity in the graph for recommendation. Path-based methods have been developed since 2013, and traditional papers call this type of method as recommendation in the HIN. 
In general, these models take advantage of the connectivity similarity of users and/or items to enhance the recommendation. To measure the connectivity similarity between entities in the graph, \textit{PathSim}~\cite{sun2011pathsim} is commonly used. It is defined as 
\begin{equation}
s_{x, y}=\frac{2 \times\left|\left\{p_{x \leadsto y}: p_{x \leadsto y} \in \mathcal{P}\right\}\right|}{\left|\left\{p_{x \leadsto x}: p_{x \leadsto x} \in \mathcal{P}\right\}\right|+\left|\left\{p_{y \leadsto y}: p_{y \leadsto y} \in \mathcal{P}\right\}\right|},
\end{equation} 
where $p_{m \leadsto n}$ is a path between the entity $m$ and $n$.
\par
One type of path-based method leverages semantic similarities of entities in different meta-paths as the graph regularization to refine the representation of users and items in the HIN. Then, $u_i$'s preference for $v_j$ can be predicted by following Equation~\ref{equation:emb-predict}, where $f(\cdot)$ refers to the inner product. 
Three types of entity similarities are commonly utilized,
\\
\textbf{$\bullet$ User-User Similarity:} the objective function for this term is 
\begin{equation}
\min _{\textbf{U}, \Theta} \sum_{l=1}^{L} \theta_{l} \sum_{i=1}^{m} \sum_{j=1}^{m} s_{i, j}^{l}\left\|\textbf{u}_{i}-\textbf{u}_{j}\right\|_{F}^{2}.
\end{equation}
where $\|\cdot\|_{F}$ denotes the matrix Frobenius norm, $\Theta = [\theta_{1}, \theta_{2},\cdots, \theta_{L}]$ denotes the weight for each meta-path, $\textbf{U} = [\textbf{u}_{1}, \textbf{u}_{2},\cdots, \textbf{u}_{m}]$ denotes latent vectors of all users, and $s_{i, j}^{l}$ denotes the similarity score of user i and j in meta-path $l$. The user-user similarity forces the embeddings of users to be close in the latent space if users share high meta-path-based similarity.
\\
\textbf{$\bullet$ Item-Item Similarity:} the objective function for this term is 
\begin{equation}
\min _{\textbf{V}, \Theta} \sum_{l=1}^{L} \theta_{l} \sum_{i=1}^{n} \sum_{j=1}^{n} s_{i, j}^{l}\left\|\textbf{v}_{i}-\textbf{v}_{j}\right\|_{F}^{2}.
\end{equation}
where $\textbf{V} = [\textbf{v}_{1}, \textbf{v}_{2},\cdots, \textbf{v}_{n}]$ denotes latent vectors of all items. Similar to the user-user similarity, the low-rank representations of items should be close if their meta-path-based similarity is high.  
\\
\textbf{$\bullet$ User-Item Similarity:} the objective function for this term is 
\begin{equation}
\min _{\textbf{U}, \textbf{V}, \Theta} \sum_{l=1}^{L} \theta_{l} \sum_{i=1}^{m} \sum_{j=1}^{n} (\textbf{u}_{i}^T\textbf{v}_{j}-s_{i, j}^{l})^2.
\end{equation}
The user-item similarity term will force the latent vector of users and items to be close to each other if their meta-path-based similarity is high. 
\par
Yu \etal~\cite{yu2013collaborative} proposed the Hete-MF, which extracts $L$ different meta-paths and calculates item-item similarity in each path. The item-item regularization is integrated with the weighted non-negative matrix factorization method~\cite{zhang2006learning} to refine low-rank representation of users and items for better recommendation.
Later, Luo \etal~\cite{luo2014hete} proposed Hete-CF to find the user's affinity to unrated items by taking the user-user similarity, item-item similarity, and user-item similarity together as regularization terms. Therefore, the Hete-CF outperforms the Hete-MF model. 

Yu \etal~\cite{yu2013recommendation} proposed HeteRec, which leverages the meta-path similarities to enrich the user-item interaction matrix $R$, so that more comprehensive representations of users and items can be extracted. HeteRec first defines $L$ different types of meta-paths that connect users and items in the HIN. The item-item similarity in each path is measured with \textit{PathSim}~\cite{sun2011pathsim}, which further forms $L$ item-item similar matrices $S^{(l)} \in \mathbb{R}^{n \times n}$, where $l = 1, 2, \cdots, L$. 
Next, $L$ diffused user preference matrices $\tilde{R}^{(q)}$ are calculated via the equation $\tilde{R}^{(l)}=R S^{(l)}$. Then
$L$ refined latent vectors of users and items in different meta-paths can be obtained via applying the non-negative matrix factorization technique~\cite{ding2008convex} on these diffused user preference matrices, 
\begin{small}
	\begin{equation}
	\begin{aligned}
	\left(\hat{\textbf{U}}^{(l)}, \hat{\textbf{V}}^{(l)}\right)=
	\operatorname{argmin}_{\textbf{U}, \textbf{V}}\left\|\tilde{R}^{(l)}-\textbf{U} ^T \textbf{V}\right\|_{F}^{2} \text { s.t. } 
	\textbf{U} \geq 0, \quad \textbf{V} \geq 0 .
	\end{aligned}
	\end{equation}
\end{small}
Finally, the recommendation can be generated by combining the user's preference on each path, with the scoring function
\begin{equation}
\hat{y}_{i, j}=\sum_{l=1}^{L} \theta_{l} \cdot \hat{\textbf{u}}_{i}^{(l)T} \hat{\textbf{v}}_{j}^{(l)},
\end{equation}
where $\theta_{l}$ is the weight for the user-item latent vector pair in the $l$-th path.
\par
Later, Yu \etal~\cite{yu2014personalized} proposed HeteRec-p, which further considers the importance of different meta-paths should vary for different users.
HeteRec-p first clusters users based on their past behaviors into $c$ groups and generates personalized recommendation with the clustering information, instead of applying a global preference model. The modified scoring function becomes 
\begin{equation}
\hat{y}_{i, j}=\sum_{k=1}^{c} \operatorname{sim}\left(\textbf{C}_{k}, \textbf{u}_{i}\right) \sum_{l=1}^{L} \theta_{l}^{k} \cdot \hat{\textbf{u}}_{i}^{(l)T} \hat{\textbf{v}}_{j}^{(l)},
\end{equation}
where $\operatorname{sim}\left(\textbf{C}_{k}, \textbf{u}_{i}\right)$ denotes the cosine similarity between user $u_i$ and the target user group $C_k$, and $\theta_{l}^{k}$ denotes the importance of meta-path $l$ for the user group $k$. 
\par
To overcome the limitation of the meta-path's representation ability, Zhao \etal~\cite{zhao2017meta} designed FMG by replacing the meta-path with the meta-graph. As a meta-graph contains richer connectivity information than a meta-path, FMG can capture the similarity between entities more accurately. Then, the model utilizes the matrix factorization (MF) to generate the latent vectors for both users and items in each meta-graph. Next, the factorization machine (FM) is applied to fuse the features of users and items across different meta-graphs for computing preference score $\hat{y}_{i, j}$. The FM considers the interaction of entities along different meta-graphs, which can further exploit connectivity patterns.
\par
The above-mentioned path-based methods only utilize the data of user's favored interacted items. Shi \etal \cite{shi2015semantic} proposed the SemRec which considers the interaction of user's favored and hated past items. This framework utilizes a weighted HIN and weighted meta-path to integrate attribute values in the link. By modeling both positive and negative preference patterns, more accurate item relations and user similarity can be depicted via these paths to propagate the real user preference.
\par
Another disadvantage of previous methods is the tedious requirement of tuning hyper-parameters, for example, the number of selected meta-paths. To lighten the burden, 
Ma \etal~\cite{ma2019jointly} proposed RuleRec to learn relations between associated items (co-buy, co-view, etc.) by exploiting the item's connectivity in an external KG. RuleRec jointly trains a rule learning module and an item recommendation module. The rule learning module first links items with associated entities in an external KG. Next, it summarizes explainable rules, which is in the form of meta-paths in the KG. The corresponding weight for each rule is further learned. Then, the item recommendation module integrates the learned rules and rule weights with the user purchase history to generate recommendations with the MF technique. Since the rules and rule weights are explicit, this model makes the recommendation process explainable.
\par
Recently, some frameworks have been proposed to learn the explicit embedding of paths that connect user-item pairs in order to directly model the user-item relations. Assume there are $K$ paths that connect $u_i$ and $v_j$ in the KG, the embedding of path $p$ is represented as $\mathbf{h}_p$. Then, the final representation of the interaction between $u_i$ and $v_j$ can be obtained via
\begin{equation}
\textbf{h} = g(\mathbf{h}_p), p = 1, 2,\cdots,K,
\label{equa:path_h}
\end{equation}
where $g(\cdot)$ is the function to summarize the information from each path embedding, which can be a max-pooling operation or weighted sum operation.   
Then, $u_i$'s preference for the $v_j$ can be modeled via 
\begin{equation}
\hat{y}_{i, j}=f(\textbf{u}_i,\textbf{v}_j,\textbf{h})
\label{equa:path_relation}
\end{equation}
where $f(\cdot)$ is the function to map the representation of the interaction between the user-item pair as well as the embedding of the user-item pair to a preference score. A common selection for $f(\cdot)$ is a fully-connected layer. 
\par
For instance, Hu \etal~\cite{hu2018leveraging} proposed MCRec, which learns the explicit representations of meta-paths to depict the interaction context of user-item pairs. For each $u_i$ and $v_j$, MCRec first uses a look up layer to embed the user-item pair. Next, it defines $L$ meta-path that connects $u_i$ and $v_j$ and samples $K$ path instances for each meta-path. These path instances are embedded with CNN to obtain the representations of each path instance $\textbf{h}_p$. Then, meta-path embeddings are calculated by applying the max-pooling operation on embeddings of path instances that belong to each type of meta-path. These meta-path embeddings are aggregated to obtain the final interaction embedding $\textbf{h}$ via an attention mechanism. The representations of the user and item also get updated via the attention mechanism with the final interaction embedding $\textbf{h}$. Finally, the preference score is calculated via Equation~\ref{equa:path_relation}, where $f(\cdot)$ is an MLP layer.
Sun \etal~\cite{Sun:2018:RKG:3240323.3240361} proposed a recurrent knowledge graph embedding (RKGE) approach that mines the path relation between user $u_i$ and item $v_j$ automatically, without manually defining meta-paths. Specifically, RKGE first enumerates user-to-item paths $\mathcal{P}(u_i, v_j)$ that connects $u_i$ and $v_j$ with different semantic relations under a sequence length constraint. Then, each path constructed by the entity embedding sequence is fed into a recurrent network to encode the entire path. Next, following Equation~\ref{equa:path_h}, final hidden states $\textbf{h}_p$ of all these paths are aggregated via the average-pooling operation to model the semantic relation $\textbf{h}$ between $u_i$ and $v_j$. Finally, the preference of $u_i$ for $v_j$ is estimated with $\textbf{h}$, and Equation~\ref{equa:path_relation} becomes $\hat{y}_{i, j}=f(\textbf{h})$, where $f(\cdot)$ is a fully-connected layer. By leveraging the information of semantic paths between entity pairs, a better representation for $u_i$ and $v_j$ will be obtained and further be integrated with the recommendation generation. 
Similarly, Wang \etal~\cite{wang2019explainable} proposed a knowledge-aware path recurrent network (KPRN) solution. KPRN constructs the extracted path sequence with both the entity embedding and the relation embedding. These paths are encoded with an LSTM layer and the preference of $u_i$ for $v_j$ in each path is predicted through fully-connected layers. By aggregating the score in each path via a weighted pooling layer, the final estimation of preference can be used for recommendation.
\par
Huang \etal~\cite{huang2019explainable} designed EIUM, which captures users' dynamic interests for sequential recommendation. The recommendation module follows the schedule in Equation \ref{equa:path_h} and \ref{equa:path_relation}. First, each path connecting the user-item pair is encoded and be aggregated to obtain the interaction embedding $\textbf{h}$ of the user-item pair $(u_i,v_j)$. The dynamic preference embedding $\textbf{p}$ is further obtained by applying the attention mechanism on the interaction sequential. The preference score can be modeled via $\hat{y}_{i, j}=f(\textbf{h},\textbf{p})$. Besides the path-based recommendation module, EIUM further integrates a multi-modal fusion constraint module. This module introduces the KG structural constraint into the framework, 
\begin{equation}
\begin{array}{ll}
{c 2 c: \textbf{e}_{h_{f_{c}}}+\textbf{r} \approx \textbf{e}_{t_{f_{c}}}}, &
{\text {s2s}: \textbf{e}_{h_{f_{s}}}+\textbf{r} \approx \textbf{e}_{t_{f_{s}}}}, \\ 
{c 2 s: \textbf{e}_{h_{f_{c}}}+\textbf{r} \approx \textbf{e}_{t_{f_{s}}}}, & 
{\text {s2c}: \textbf{e}_{h_{f_{s}}}+\textbf{r} \approx \textbf{e}_{t_{f_{c}}}},
\end{array}
\end{equation}
where $(e_h,r,e_t) \in \mathcal{G}$, $f_c$ denotes the content feature (textual, visual), and $f_s$ denotes the structural feature. The loss function of this module is 
\begin{equation}
\begin{aligned} 
\mathcal{L}_{KG} &=\mathcal{L}_{c 2 c}+\mathcal{L}_{s 2 s}+\mathcal{L}_{c 2 s}+\mathcal{L}_{s 2 c} \\ &=\frac{1}{4} \sum_{i}\|h+r-t\|, i \in\{c 2 c, s 2 s, c 2 s, s 2 c\}. 
\end{aligned}
\end{equation}
This term can refine features of entities under the structural constraint of the KG. In this way, more accurate recommendation can be generated.  
\par
Recently, Xian \etal~\cite{xian2019reinforcement} proposed Policy-Guided Path Reasoning (PGPR) to use reinforcement learning (RL) to search for reasonable paths between user-item pairs. They formulated the recommendation problem as a Markov decision process to find a reasonable path connecting the user-item pair in the KG. They trained an agent to sample paths between users and items by carefully designing the path searching algorithm, the transition strategy, terminal conditions, and RL rewards. In the prediction phase, PGPR can generate recommended items for users with specific paths to interpret the reasoning process. 
Later, Song \etal~\cite{song2019explainable} proposed a similar model, EKar*, which adopts the RL technique in generating recommendation as well.
\\
\textbf{Summary for Path-based Methods.} 
Path-based methods generate recommendations based on user-item graphs, and such methods have also been called HIN-based recommendation in the past. Traditional path-based methods~\cite{luo2014hete,yu2013collaborative,yu2013recommendation,yu2014personalized,zhao2017meta,shi2015semantic,shi2018heterogeneous} generally integrate MF with extracted meta-paths in HINs. These methods utilize path connectivity to regularize or enrich the user and/or item representation. The disadvantage of these methods is that they commonly need domain knowledge to define the type and number of meta-paths. RuleRec~\cite{ma2019jointly} tries to overcome the limitation by exploiting rules in an external KG in an automatic fashion. With the development of deep learning techniques, different models~\cite{hu2018leveraging,Sun:2018:RKG:3240323.3240361,wang2019explainable,xian2019reinforcement,song2019explainable,huang2019explainable} have been proposed to encode the path embedding explicitly. Recommendation can be generated with the path embeddings, or by discovering the most salient paths that connect user-item pairs. 
\par
Path-based methods naturally bring interpretability into the recommendation process. For traditional path-based methods, the motivation is to match the similarity of the item or user on the meta-path level. The recommendation results can find a reference from the pre-defined meta-paths. RuleRec utilize an external KG to generate rules for recommendation. Since the rule and corresponding weight are explicit, the reason for recommendation is also available to users. More recent works take advantage of deep learning models to mine salient paths for a user-item pair automatically, which reflects the recommendation process in the graph.

\subsection{Unified Methods}
As discussed in Section~\ref{section4-1:embedding} and Section~\ref{section4-2:path}, embedding-based methods leverage the semantic representation of users/items in the KG for recommendation, while path-based methods use the semantic connectivity information, and both approaches utilize only one aspect of information in the graph. To fully exploit the information in the KG for better recommendations, unified methods which integrate both the semantic representation of entities and relations, and the connectivity information have been proposed. 
The unified method is based on the idea of embedding propagation. 
These methods refine the entity representation with the guidance of the connective structure in the KG. After obtaining the enriched representations of user $u_i$ and/or the potential item $v_j$, the user's preference can be predicted with Equation~\ref{equation:emb-predict}.
\par
The first group of works refine the user's representation from their interaction history. These works first extract multi-hop ripple sets $\mathcal{S}_{u_i}^{k} (k = 1,2, \cdots, H)$ (defined in Section~\ref{section3:overview}), where $\mathcal{S}_{u_i}^{1}$ is the triple set $(e_h, r, e_t)$ in the graph with the head entities being the user $u_i$'s engaged items. The general idea of this method is to learn the user embedding by utilizing the embeddings of past interacted items as well as multi-hop neighbors of these interacted items. The process of learning user representation $\mathbf{u_i}$ can be written in a general form as
\begin{equation}
\mathbf{u_i} = g_{u}\left(\left\{\mathcal{S}_{u_i}^{k}\right\}_{k=1}^{H}\right),
\end{equation}
where $g_u(\cdot)$ is a function to concatenate embeddings of multi-hop entities with bias. Since the propagation starts from the user's engaged items, this process can be regarded as propagating the user's preference in the graph. 
\par
Wang \etal~\cite{wang2018ripplenet} proposed RippleNet, which is the first work to introduce the concept of preference propagation. 
Specifically, RippleNet first assigns entities in the KG with initial embeddings. Then it samples ripple sets $\mathcal{S}_{u_i}^{k} (k=1,2, \cdots, H)$ from the KG.
To refine the user representation, the aggregation process can be illustrated as follows. Starting from $S_{u_i}^{1}$, every head entity interacts with the embedding of the candidate item $v_j$ in turn via
\begin{equation}
p_{i}=
\frac{\exp \left(\mathbf{v}_j^{{T}} \mathbf{R}_{i} \mathbf{e}_{{h}_{i}}\right)}{\sum_{(e_{h_k}, r_k, e_{t_k}) \in \mathcal{S}_{u_i}^{1}} \exp \left(\mathbf{v}_j^{{T}} \mathbf{R}_k \mathbf{e}_{{h}_k}\right)},
\label{equa:out1}
\end{equation}
where $\mathbf{R}_{i} \in \mathbb{R}^{d \times d}$ represents the embedding of relation $r_i$, and $\mathbf{e}_{{h}_{i}} \in \mathbb{R}^{d}$ is the embedding of head entity in the ripple set. During this process, the similarities of the candidate item $v_j$ and head entities are calculated in the relation space. Then, the user's $1$-order response of historical interaction can be calculated via  
\begin{equation}
\mathbf{o}_{u_i}^{1}=\sum_{\left(e_{h_{i}}, r_{i}, e_{t_{i}}\right) \in \mathcal{S}_{u_i}^{1}} p_{i} \mathbf{e}_{{t}_{i}},
\label{equa:out2}
\end{equation}
where $\mathbf{e}_{t_i}$ represents the embedding of the tail entity in the ripple set. The user's $h$-order $(h=2,3,\cdots,H)$ response $\mathbf{o}_{u_i}^{h}$ can be obtained by replacing $\mathbf{v}_j$ with the $(h-1)$-order response $\mathbf{o}_{u}^{h-1}$ in Equation \ref{equa:out1}, then interacting with head entities in $h$-hop ripple set $\mathcal{S}_{u}^{h}$ iteratively. The final representation of $u_i$ can be obtained with the equation of $\mathbf{u}_i=\mathbf{o}_{u_i}^{1}+\mathbf{o}_{u_i}^{2}+\cdots+\mathbf{o}_{u_i}^{H}$. 
Finally, the preference score can be generated with 
\begin{equation}
\hat{y}_{i, j}=\sigma\left(\mathbf{u_i}^{{T}} \mathbf{v_j}\right),
\end{equation}
where $\sigma(x)$ is the sigmoid function. In this way, RippleNet propagates the user's preference from historical interests along the path in the KG.
\par
Similar to RippleNet, Tang \etal~\cite{tang2019akupm} proposed AKUPM, which models users with their click history. AKUPM first applies TransR for the entity representation. During each propagation process, AKUPM learns the relations between entities with a self-attention layer and propagates the user's preference toward different entities with bias. 
Finally, embeddings from different-order neighbors of interacted items are aggregated with the self-attention mechanism to obtain the final user representation. 
Later, Li \etal~\cite{li2019unifying} extended the AKUPM and designed RCoLM. RCoLM jointly trains the KG completion module and the recommendation module, where AKUPM serves as the backbone. With the assumption that an item should have the same latent representation in the two modules, RCoLm unifies two modules and facilitates their mutual enhancement. Thus, RCoLM outperforms the AKUPM model.
\par
The second group of works focus on refining the item representation $\mathbf{v}_j$ by aggregating embeddings of an item's multi-hop neighbors $\mathcal{N}_v^{k} (k=1,2,\cdots,H)$. A general description for this process is 
\begin{equation}
\mathbf{v}_j = g_{v}\left(\left\{\mathcal{S}_{v_j}^{k}\right\}_{k=1}^{H}\right),
\end{equation}
where $\mathcal{S}_{v_j}^{k}$ is the ripple set of candidate item $v_j$, and $g_v(\cdot)$ is the function to concatenate embeddings of multi-hop neighbors. There are two steps to concatenate the embeddings of multi-hop neighbors. The first step is to learn a representation of candidate item $v_j$'s $k$-hop neighbors, 
\begin{equation}
\mathbf{e}_{\mathcal{S}_{v_j}^{k}} = \sum_{(e_h,r,e_t) \in \mathcal{S}_{v_j}^{k}} \alpha_{(e_h,r,e_t)} \mathbf{e}_{t}, 
\end{equation}
where $\alpha_{(e_h,r,e_t)}$ denotes the importance of different neighbors. Then for $e_h \in \mathcal{S}_{v_j}^{k}$, the representation can be updated by 
\begin{equation}
\mathbf{e}_{h}=\operatorname{agg}\left(\mathbf{e}_{h}, \mathbf{e}_{\mathcal{S}_{v_j}^{k}}\right),
\label{equa:agg}
\end{equation}
where $\operatorname{agg}$ is the aggregation operator. During this process, the information of $k$-hop neighbors is aggregated with that of $(k-1)$-hop neighbors.   
Four types of aggregators are commonly used:
\\
\textbf{$\bullet$ Sum Aggregator.} The sum aggregator sums two representations, followed by a nonlinear transformation.
\begin{equation}
\operatorname{agg}_{\operatorname{sum}}=\Phi\left(\mathbf{W} \cdot\left(\mathbf{e}_h+\mathbf{e}_{\mathcal{S}_{v_j}^{k}}\right)+\mathbf{b}\right).
\end{equation}
\\
\textbf{$\bullet$ Concat Aggregator.} The concat aggregator concatenates two representations, then applies a nonlinear transformation.
\begin{equation}
\operatorname{agg}_{\text {concat }}=\Phi\left(\mathbf{W} \cdot\left(\mathbf{e}_h\oplus\mathbf{e}_{\mathcal{S}_{v_j}^{k}}\right)+\mathbf{b}\right).
\end{equation}
\\
\textbf{$\bullet$ Neighbor Aggregator.} The neighbor aggregator directly replaces the representation of an entity with representations from neighbors. 
\begin{equation}
\operatorname{agg}_{\text {neighbor}}=\Phi\left(\mathbf{W} \cdot \mathbf{e}_{\mathcal{S}_{v_j}^{k}}+\mathbf{b}\right).
\end{equation}
\\
\textbf{$\bullet$ Bi-Interaction Aggregator.} The bi-interaction aggregator considers both the sum and the element-wise product relations between entities. The second term allows more information to be passed from similar entities. 
\begin{equation}
\begin{aligned}
\operatorname{agg}_{\text {Bi-Interaction}}=
&\Phi\left(\mathbf{W} \cdot\left(\mathbf{e}_h+\mathbf{e}_{\mathcal{S}_{v_j}^{k}}\right)+\mathbf{b} \right) +\\
&\Phi\left(\mathbf{W} \cdot\left(\mathbf{e}_h \odot \mathbf{e}_{\mathcal{S}_{v_j}^{k}}\right)+\mathbf{b} \right).
\end{aligned}
\end{equation}
\par
Wang \etal~\cite{Wang:2019:KGC:3308558.3313417} proposed KGCN which models the final representation of a candidate item $v_j$ by aggregating the embedding of entities in the KG from distant neighbors of $v_j$ to $v_j$ itself. KGCN first samples neighbors of the candidate item $v_j$ in the KG, and it iteratively samples neighbors with a fixed number for each entity. Starting from the $H$-hop neighbors, it updates the representation of inner entities by replacing $k=H,H-1,\cdots,1$ iteratively in Equation \ref{equa:agg}.
During the aggregation process, the information of multi-hop neighbors can be propagated to the candidate item $v_j$ inwardly. After this feature propagation process, the final representation of item $v_j$ is a mixture of its initial representation and information from multi-hop neighbors. RippleNet and KGCN are two similar frameworks, the former models users by propagating the user's preference from historical interests outwardly, while the latter learns item representations from distant neighbors inwardly. Moreover, KGCN leverages the idea of GCN by sampling a fixed number of neighbors as the receptive field, which makes the learning process highly efficient and scalable.
Recently, Wang \etal~\cite{Wang:2019:KGN:3292500.3330836} proposed a follow-up approach, KGCN-LS, which further adds a label smoothness (LS) mechanism on the KGCN model. The LS mechanism takes the information of user interaction and propagates the user interaction labels on the KG, which is able to guide the learning process and obtain a comprehensive representation for the candidate item $v_j$.
\par
RippleNet and its extension focus on using the embedding propagation mechanism on the item KG.   
Recently, some papers have explored the propagation mechanism in the user-item graph. 
Wang \etal~\cite{Wang:2019:KKG:3292500.3330989} proposed KGAT, which directly models the high order relations between users and items with embedding propagation. KGAT first applies TransR to obtain the initial representation for entities. Then, it runs the entity propagation from the entity itself outwardly. During the outward propagation process, information from the entity $e_i$ will be interacted with the multi-hop neighbors iteratively. 
The Equation \ref{equa:agg} can be modified as   
\begin{equation}
\mathbf{e}_{i}^{k+1}=\operatorname{agg}\left(\mathbf{e}_{i}^{k}, \mathbf{e}_{\mathcal{S}_{e_i}^{k+1}}\right), k=0,1,\cdots,H-1,
\end{equation}
where $\mathbf{e_i}^{0}$ represents the initial presentation of the entity, and $\mathbf{e_i}^{k}$ contains the connectivity information from $k$-hop neighbors. These $H$ embeddings $\mathbf{e_i}^{k}$ are aggregated with bias to form the final representation $\mathbf{e_i}^{*}$. In this way, both the user representation and the item representation can be enriched with corresponding neighbors. The user preference is modeled via $\hat{y}_{u, v}=\mathbf{e}_{u}^{* T} \mathbf{e}_{v}^{*}$, where $\mathbf{e}_{u}^{*}$ and $\mathbf{e}_{v}^{*}$ stands for the final representation of the user $u$ and item $v$, respectively.
\par
Qu \etal~\cite{qu2019end} proposed KNI, which further considers the interaction between item-side neighbors and user-side neighbors, so that the refinement process of user embeddings and item embeddings are not separated.  
Zhao \etal~\cite{zhao2019intentgc} proposed IntentGC, which exploits rich user-related behaviors in the graph for better recommendation. They also designed a faster graph convolutional network to guarantee the scalability of IntentGC.
Recently, Sha \etal~\cite{sha2019attentive} proposed AKGE, which learns the representation of user $u_i$ and candidate item $v_j$ by propagating information in a subgraph of this user-item pair. AKGE first pre-trains the embeddings of entities in the graph with TransR, then samples several paths connecting $u_i$ and $v_j$ based on the pairwise distance in these paths, which forms a subgraph for $u_i$ and $v_j$. Next, AKGE uses an attention-based GNN in this subgraph to propagate the information from neighbors for the final representation of this user-item pair. The construction of the subgraph filters out less related entities in the graph, facilitating mining high-order user-item relations for recommendation.
\\
\textbf{Summary for Unified Methods.}
Unified methods benefit from both the semantic embedding of the KG and semantic path patterns. These methods leverage the idea of embedding propagation to refine the representation of the item or user with multi-hop neighbors in the KG. These works generally adopt a GNN-based architecture that naturally fits the process of embedding propagation, and such methods have been a new research trend since the RippleNet~\cite{wang2018ripplenet} was proposed in 2018. 
Unified methods inherit interpretability from path-based methods. The propagation process can be treated as discovering user's preference patterns in the KG, which is similar to finding connectivity patterns in path-based methods. 
\subsection{Summary}
Embedding-based methods preprocess the KG, either item graph or user-item graph, with KGE methods to obtain the embedding of entities and relations, which is further integrated into the recommendation framework. However, the informative connectivity patterns in the graph are ignored in this approach and few works can provide the recommendation results with reasons. Path-based methods utilize the user-item graph to discover path-level similarity for items, either by predefining meta-paths or mining connective patterns automatically. The path-based approach can also provide users with an explanation for the result. A recent research trend is to unify the embedding-based method and the path-based method to fully exploit information from both sides. Moreover, unified methods also have the ability to explain the recommendation process.

\section{Datasets of Recommender Systems with Knowledge Graph}
Besides the benefit of accuracy and interpretability, another advantage of KG-based recommendation is that this type of side information can be naturally incorporated into recommender systems for different applications. To show the effectiveness of the KG as side information, KG-based recommender systems have been evaluated on datasets under different scenarios. In this section, we categorized these works based on the dataset and illustrate the difference among these scenarios. The contributions of this section are two-fold. First, we provide an overview of datasets used under various scenarios. Second, we illustrate how knowledge graphs are constructed for different recommendation tasks. This section can help researchers find suitable datasets to test their recommender systems. 

\begin{table*}[!]
	\centering
	\caption{\label{applications} A collection of datasets for different application scenarios and corresponding papers.}
	\begin{tabular}{lll}
		\hline
		\textbf{Scenario} & \textbf{Dataset}       & \textbf{Paper} \\ \hline
		Movie             & MovieLens-100K           &\cite{yu2013collaborative,yu2013recommendation,yu2014personalized,catherine2016personalized,Xin:2019:RCF:3331184.3331188,hu2018leveraging}\\
		& MovieLens-1M            & \cite{Zhang:2016:CKB:2939672.2939673,palumbo2017entity2rec,huang2018improving,Cao:2019:UKG:3308558.3313705,Wang:2019:MFL:3308558.3313411,Sun:2018:RKG:3240323.3240361,wang2019explainable,song2019explainable,wang2018ripplenet,tang2019akupm,qu2019end,sha2019attentive,li2019unifying}            \\
		& MovieLens-20M                    & \cite{huang2018improving,huang2019explainable,wang2019exploring,Wang:2019:KGC:3308558.3313417,Wang:2019:KGN:3292500.3330836,qu2019end}            \\
		& DoubanMovie         & \cite{yang2018knowledge,shi2015semantic,shi2018heterogeneous}          \\
		Book              & DBbook2014          & \cite{Cao:2019:UKG:3308558.3313705,song2019explainable}           \\
		& Book-Crossing       & \cite{Wang:2019:MFL:3308558.3313411,wang2018ripplenet,wang2019exploring,Wang:2019:KGC:3308558.3313417,Wang:2019:KGN:3292500.3330836,tang2019akupm,qu2019end,li2019unifying}            \\
		& Amazon-Book         & \cite{huang2018improving,Wang:2019:KKG:3292500.3330989,qu2019end}            \\
		& IntentBooks         & \cite{Zhang:2016:CKB:2939672.2939673}            \\
		& DoubanBook          & \cite{shi2018heterogeneous}          \\
		News              & Bing-News           & \cite{Wang:2018:DDK:3178876.3186175,Wang:2019:MFL:3308558.3313411,wang2018ripplenet,wang2019exploring}            \\
		Product          & Amazon Product data & \cite{zhao2017meta,ma2019jointly,zhang2018learning,xian2019reinforcement,ai2018learning,zhao2019intentgc}            \\
		& Alibaba Taobao      & \cite{ye2019bayes,zhao2019intentgc}            \\
		POI               & Yelp challenge      & \cite{shi2015semantic,zhao2017meta,hu2018leveraging,Sun:2018:RKG:3240323.3240361,shi2018heterogeneous,catherine2016personalized,yu2013recommendation,yu2014personalized,sha2019attentive,Wang:2019:KKG:3292500.3330989}         \\
		& Dianping-Food       & \cite{Wang:2019:KGN:3292500.3330836}        \\
		& CEM                 & \cite{dadoun2019location}           \\
		Music             & Last.FM             & \cite{huang2018improving,Wang:2019:MFL:3308558.3313411,hu2018leveraging,song2019explainable,Wang:2019:KGC:3308558.3313417,Wang:2019:KGN:3292500.3330836,Wang:2019:KKG:3292500.3330989,sha2019attentive}            \\
		& KKBox               & \cite{Xin:2019:RCF:3331184.3331188,wang2019explainable}            \\
		Social Platform            & Weibo               & \cite{wang2018shine}          \\
		& DBLP                & \cite{luo2014hete}        \\
		& MeetUp              & \cite{luo2014hete}        \\ \hline
	\end{tabular}
\end{table*}

We group KG based recommender systems according to the datasets which are summarized in Table \ref{applications}. Generally, these works can be categorized into seven application scenarios and we will illustrate how different works construct the KG with each dataset.
\\
\textbf{$\bullet$ Movie.} In this task, the recommender system needs to infer the user's preference based on movies watched in the past. Two datasets are most commonly used: MovieLens~\cite{movielens1998} and DoubanMovie. MovieLens maintains a set of datasets collected from the MovieLens website~\cite{movielensorg1997}, among which three stable benchmark datasets with different rating numbers, MovieLens-100K, MovieLens-1M, and MovieLens-20M are most commonly used. Each dataset contains ratings, the movie's attributes and tags.  
DoubanMovie is crawled from Douban~\cite{doubanmovie2005}, a popular Chinese social media network. The dataset includes the social relation among users and the attributes of users and movies.
\par 
There are different ways to construct the movie-related KG for recommendation. 
Some papers~\cite{Zhang:2016:CKB:2939672.2939673, yang2018knowledge, Xin:2019:RCF:3331184.3331188,huang2018improving,Cao:2019:UKG:3308558.3313705,Wang:2019:MFL:3308558.3313411,wang2018ripplenet,li2019unifying,wang2019exploring,Wang:2019:KGC:3308558.3313417,Wang:2019:KGN:3292500.3330836,tang2019akupm,qu2019end} construct the movie-centric item graph to enrich the information of movies by extracting movies and related attributes from Satori, DBpedia, Freebase, CN-DBPedia, or IMDB~\cite{imdb1990}. 
In this way, movies are connected via attributes, including genres, countries, actors, directors, etc. This item graph serves as side information to facilitate the collaborative filtering module.
Another approach is to directly take the user's rating as one type of relation and introduce the user to the graph. Some papers~\cite{shi2015semantic,hu2018leveraging,shi2018heterogeneous} build the user-item graph by directly leveraging the interaction data and attributes of movies inside the MovieLens dataset or the DoubanMovie dataset, while others~\cite{yu2013collaborative,yu2013recommendation,yu2014personalized,catherine2016personalized,palumbo2017entity2rec,Sun:2018:RKG:3240323.3240361,wang2019explainable,song2019explainable,huang2019explainable,sha2019attentive} still utilize external database to enrich the movie-side information.  
\\
\textbf{$\bullet$ Book.} Book recommendation is another popular task. There are five commonly used datasets: Book-Crossing~\cite{bookcrossing2004}, Amazon-Book~\cite{mcauley2015image}, DoubanBook, DBbook2014, and IntentBooks~\cite{uyar2015evaluating}.
Book-Crossing, DBbook2014, IntentBooks, and Amazon-Book contain binary feedback between users and books, and the KG for each dataset is built by mapping books to corresponding entities in Satori~\cite{Zhang:2016:CKB:2939672.2939673,Wang:2019:MFL:3308558.3313411,wang2018ripplenet,li2019unifying,wang2019exploring,Wang:2019:KGC:3308558.3313417,Wang:2019:KGN:3292500.3330836,tang2019akupm,qu2019end}, DBpedia~\cite{Cao:2019:UKG:3308558.3313705,song2019explainable}, or Freebase~\cite{huang2018improving,Wang:2019:KKG:3292500.3330989,qu2019end}.
The DoubanBook dataset is crawled from Douban~\cite{doubanbook2005}, which contains both the user-item interaction data and books attributes, such as information about the author, publisher, and the year of publication. This work~\cite{shi2018heterogeneous} builds the user-item graph by utilizing this knowledge in the DoubanBook dataset without the assistance of an external KG.
%
\\
\textbf{$\bullet$ Music.} Last.FM~\cite{schedl2016lfm} is the most popular dataset for music recommendation. The dataset contains information about users and their music listening records from the Last.fm online music system~\cite{lastfm2002}. Some papers~\cite{huang2018improving,Wang:2019:MFL:3308558.3313411,Wang:2019:KGC:3308558.3313417,Wang:2019:KKG:3292500.3330989,Wang:2019:KGN:3292500.3330836} construct the item graph by extracting music-related subgraphs from Freebase or Satori. Some papers~\cite{song2019explainable,sha2019attentive} build the user-item graph with knowledge from Freebase or Satori, while this paper~\cite{hu2018leveraging} build the user-item graph from the Last.FM dataset directly.
Another popular dataset is the KKBox dataset, which was released by the WSDM Cup 2018 Challenge~\cite{kkbox2018}. This dataset contains both the user-item interaction data and the description of the music. Paper~\cite{Xin:2019:RCF:3331184.3331188} builds the item graph and \cite{wang2019explainable} builds the user-item graph from this dataset without leveraging any external databases.
\\
\textbf{$\bullet$ Product.} The most popular dataset for the product recommendation task is the Amazon Product dataset~\cite{mcauley2015image}. This dataset includes multiple types of item and user information, such as interaction records, user  reviews, product categories, product descriptions, and user behaviors. These works \cite{ai2018learning,zhang2018learning,zhao2017meta,xian2019reinforcement,zhao2019intentgc} build a user-item graph with this dataset alone, and~\cite{ma2019jointly} build the item graph by enriching the item information with the external Freebase database. There are also some papers~\cite{ye2019bayes,zhao2019intentgc} use the data provided by Alibaba Taobao.
\\
\textbf{$\bullet$ POI.} Point of Interest (POI) recommendation is the recommendation of new businesses and activities (restaurants, museums, parks, cities, etc.) to users based on their historical check-in data. The most popular dataset is the Yelp Challenge~\cite{yelp2013}, which contains the information of businesses, users, check-ins, and reviews. These papers~\cite{yu2013recommendation,yu2014personalized,shi2015semantic,catherine2016personalized,zhao2017meta,hu2018leveraging,Sun:2018:RKG:3240323.3240361,shi2018heterogeneous,sha2019attentive} build a user-item graph with the data of check-ins, reviews and the attributes in the dataset, while \cite{Wang:2019:KKG:3292500.3330989} construct the item graph. Paper~\cite{dadoun2019location} utilizes the CEM dataset\footnote{an Amadeus database containing bookings over a dozen of airlines} to recommend next trip. Another work~\cite{Wang:2019:KGN:3292500.3330836} uses the Dianping-Food dataset, which is provided by Dianping.com~\cite{dianping2009} for restaurant recommendation.
\\
\textbf{$\bullet$ News.} News recommendation is challenging~\cite{Wang:2018:DDK:3178876.3186175} because the news itself is time-sensitive, and the content is highly condensed, which requires commonsense to understand. Moreover, people are topic-sensitive in choosing news to read and may prefer news from various domains. Traditional news recommendation models fail to discover the high level connection among the news. 
Therefore, KGs are introduced into this scenario~\cite{Wang:2018:DDK:3178876.3186175,Wang:2019:MFL:3308558.3313411,wang2018ripplenet,wang2019exploring} to find the logical relations between different news and improve the precision of recommendation. 
The most popular dataset is Bing-News, collected from server logs of Bing News~\cite{bingnews2009}, which contains the user click information, news title, etc.
To build a KG for news recommendation, the first step is to extract entities in the title. Then, subgraphs are constructed by extracting neighbors of these entities in Satori. 
\\
\textbf{$\bullet$ Social Platform.} This task is to recommend potentially interested people or meetings to users in the community. One application is to recommend unfollowed celebrities to users on the social platform Weibo~\cite{weibo2009} with the collected Weibo tweets data~\cite{wang2018shine}. Despite the user-item graph to represent sentiment links between users and celebrities, an item graph with knowledge extracted from the Satori is built to enrich the information of celebrities. Another application is to recommend offline meetings for users on a social website, MeetUp~\cite{meetup2002}, with data on that platform. The last application lies in the academic domain, to recommend conferences to researchers with the DBLP data~\cite{dblp2013}.


\section{Future Directions}
In the above sections, we have demonstrated the advantage of KG-based recommender systems from the aspects of more accurate recommendation and explainability. Although many novel models have been proposed to utilize the KG as side information for recommendation, some further opportunities still exist. In this section, we outline and discuss some prospective research directions.
\\
\textbf{$\bullet$ Dynamic Recommendation.} Although KG-based recommender systems with GNN or GCN architectures have achieved good performance, the training process is time-consuming. Thus such models can be regarded as static preference recommendation. However, in some scenarios, such as online shopping, news recommendation, Twitter, and forums, a user's interest can be influenced by social events or friends very quickly. In this case, recommendation with a static preference modeling may not be enough to understand real-time interests. In order to capture dynamic preference, leveraging the dynamic graph network can be a solution. Recently, Song \etal~\cite{song2019session} designed a dynamic-graph-attention network to capture the user's rapidly-changing interests by incorporating long term and short term interests from friends. It is natural to integrate other types of side information and build a KG for dynamic recommendation by following such an approach.
\\
\textbf{$\bullet$ Multi-task Learning.} KG-based recommender systems can be naturally regarded as link prediction in the graph. Therefore, considering the nature of the KG has the potential to improve the performance of graph-based recommendation. For example, there may exist missing facts in the KG, which leads to missing relations or entities. However, the user's preference may be ignored because these facts are missing, which can deteriorate the recommendation results. \cite{Cao:2019:UKG:3308558.3313705, li2019unifying} have shown it is effective to jointly train the KG completion module and recommendation module for better recommendation. Other works have utilized multi-task learning by jointly training the recommendation module with the KGE task~\cite{Wang:2019:MFL:3308558.3313411} and item relation regulation task~\cite{Xin:2019:RCF:3331184.3331188}. 
It would be interesting to exploit transferring knowledge from other KG-related tasks, such as entity classification and resolution, for better recommendation performance.  
\\
\textbf{$\bullet$ Cross-Domain Recommendation.} Recently, works on cross-domain recommendation have appeared. The motivation is that interaction data is not equal across domains. For example, on the Amazon platform, book ratings are denser than other domains. With the transfer learning technique, interaction data from the source domain with relatively rich data can be shared for better recommendation in the target domains.
Zhang~\etal~\cite{zhang2018cross} proposed a matrix-based method for cross-domain recommendation. Later, Zhao~\etal~\cite{zhao2019cross} introduced PPGN, which puts users and products from different domains in one graph, and leverages the user-item interaction graph for cross-domain recommendation. Although PPGN outperforms SOTA significantly, the user-item graph contains only interaction relations, and does not consider other relationships among users and items. It could be promising to follow works in this survey, by incorporating different types of user and item side information in the user-item interaction graph for better cross-domain recommendation performance.
\\
\textbf{$\bullet$ Knowledge Enhanced Language Representation.} 
To improve the performance of various natural language processing tasks, there is a trend to integrate external knowledge into the language representation model. The knowledge representation and the text representation can be refined mutually. For example, Chen \etal~\cite{chen2019deep} proposed the STCKA for short text classification, which utilizes the prior knowledge from KGs, such as YAGO, to enrich the semantic representation of short texts. Zhang \etal~\cite{zhang2019ernie:} proposed the ERNIE, which incorporates knowledge from Wikidata to enhance the language representation, and such an approach has proven to be effective in the task of relation classification.
Although the DKN model~\cite{Wang:2018:DDK:3178876.3186175} utilizes both the text embedding and the entity embedding in the news, these two types of embeddings are simply concatenated to obtain the final representation of news, instead of considering the information fusion between two vectors.   
Therefore, it is promising to apply the strategy of knowledge-enhanced text representation in the news recommendation task and other text-based recommendation tasks for better representation learning to achieve more accurate recommendation results. 
\\
\textbf{$\bullet$ Knowledge Graph Embedding Method.} There are two types of KGE methods, translation distance models and semantic matching models, based on the different constraints. 
In this survey, these two types of KGE methods are used in all three kinds of KG-based recommender systems and recommendation tasks. However, there is no comprehensive work to suggest under which circumstances, including data sources, recommendation scenarios, and model architectures, should a specific KGE method be adopted. Therefore, another research direction lies in comparing the advantages of different KGE methods under various conditions.
\\
\textbf{$\bullet$ User Side Information.} Currently, most KG-based recommender systems build the graph by incorporating item side information, while few models consider user side information. However, user side information, such as the user network, and user's demographic information, can also be naturally integrated into the framework of current KG-based recommender systems. Recently, Fan \etal~\cite{fan2019graph} used the GNN to represent a user-user social network and a user-item interaction graph separately, which outperforms traditional CF-based recommender systems with user social information. A recent paper in our survey~\cite{sha2019attentive} integrated user relations into the graph and showed the effectiveness of this strategy. Therefore, considering user side information in the KG could be another research direction. 



\section{Conclusion}
In this survey paper, we investigate KG-based recommender systems and summarize the recent efforts in this domain. This survey illustrates how different approaches utilize the KG as side information to improve the recommendation result as well as providing interpretability in the recommendation process. Moreover, an introduction to datasets used in different scenarios is provided. Finally, future research directions are identified, hoping to promote development in this field. KG-based recommender systems are promising for accurate recommendation and explainable recommendation, benefitting from the fruitful information contained in the KGs. We hope this survey paper can help readers better understand work in this area.

\ifCLASSOPTIONcompsoc
  \section*{Acknowledgments}
\else
  \section*{Acknowledgment}
\fi

The research work supported by the National Key Research and Development Program of China under Grant No. 2018YFB1004300, the National Natural Science Foundation of China under Grant No. U1836206, U1811461, 61773361, the Project of Youth Innovation Promotion Association CAS under Grant No. 2017146.




\bibliographystyle{IEEEtran}
\bibliography{KG_RS}

%

\end{document}